\documentclass{article}

\usepackage[english]{babel}

\usepackage[letterpaper,top=2cm,bottom=2cm,left=3cm,right=3cm,marginparwidth=1.75cm]{geometry}

\usepackage{amsmath,amsfonts,bm}

\def\1{\bm{1}}

\DeclareMathAlphabet{\mathsfit}{\encodingdefault}{\sfdefault}{m}{sl}
\SetMathAlphabet{\mathsfit}{bold}{\encodingdefault}{\sfdefault}{bx}{n}

\usepackage{amsmath,amssymb}
\usepackage{graphicx}
\usepackage{algorithm,algorithmic}
\usepackage{color}
\usepackage{url}
\usepackage{cite}

\usepackage{booktabs}
\usepackage{comment}

\usepackage[pagebackref,breaklinks,colorlinks]{hyperref}

\newcommand{\ten}[1]{ \boldsymbol{\mathcal #1}}
\newcommand{\bbR}[1]{\mathbb{R}^{#1}}

\newcommand{\eg}{{\em e.g. }}
\newcommand{\ie}{{\em i.e. }}

\usepackage[capitalize]{cleveref}
\crefname{section}{Sec.}{Secs.}
\Crefname{section}{Section}{Sections}
\Crefname{table}{Table}{Tables}
\crefname{table}{Tab.}{Tabs.}

\usepackage{lipsum}
\newcommand\blfootnote[1]{%
\begingroup
\renewcommand\thefootnote{}\footnote{#1}%
\addtocounter{footnote}{-1}%
\endgroup
}

\title{{\Large Soft Smoothness for Audio Inpainting \\ Using a Latent Matrix Model in Delay-embedded Space}}
\author{Tatsuya Yokota$^{\dagger,\ddagger}$\\
  {\small $^\dagger$Nagoya Institute of Technology, Aichi, Japan} \\
{\small $^{\ddagger}$RIKEN Center for Advanced Intelligence Project, Tokyo, Japan}}

\date{ }

\begin{document}
\maketitle

\begin{abstract}
  Here, we propose a new reconstruction method of smooth time-series signals.
  A key concept of this study is not considering the model in signal space, but in delay-embedded space.
  In other words, we indirectly represent a time-series signal as an output of inverse delay-embedding of a matrix, and the matrix is constrained.
  Based on the model under inverse delay-embedding, we propose to constrain the matrix to be rank-1 with smooth factor vectors.
  The proposed model is closely related to the convolutional model, and quadratic variation (QV) regularization.
  Especially, the proposed method can be characterized as a generalization of QV regularization.
  In addition, we show that the proposed method provides the softer smoothness than QV regularization.
  Experiments of audio inpainting and declipping are conducted to show its advantages in comparison with several existing interpolation methods and sparse modeling.
\blfootnote{This work was supported by Japan Science and Technology Agency (JST) ACT-I under Grant JPMJPR18UU.}
\end{abstract}

\section{Introduction}
Mathematical models that assume smoothness play an important role in a wide range of fields such as signal processing and pattern recognition.
Smooth mathematical models are directly useful for noise reduction and interpolation of corrupted signals, and have many applications in time-series restoration \cite{selesnick2012total}, image restoration \cite{vogel1998fast,guichard1998total,osher2005iterative,goldfarb2005second,zhu2008efficient,wang2008new,oliveira2009adaptive,yokota2017efficient}, color-image restoration \cite{han2014linear,ono2014decorrelated,guo2015generalized,ono2016vectorial,yokota2017simultaneous,yokota2019simultaneous}, MR image reconstruction \cite{shi2013low,li2017low,kawamura2018super}, dynamic PET reconstruction \cite{kawai2017robust,kawai2018simultaneous,yokota2019dynamic}, and hyper spectral image restoration \cite{iordache2012total,yuan2012hyperspectral}.

There are two approaches for constructing smooth signals, a generative approach and a constrained approach \cite{yokota2021low}.
The generative approach defines a parametric model of smooth signals/functions and optimizes its parameters to the given data.
This includes polynomial models \cite{de1978practical,fan1996local}, tensor factorization \cite{reis2002parafac}, and neural networks \cite{bishop1995neural}.
In the formula, it can be simply written as
\begin{align}
    \mathop{\text{minimize}}_{\theta} D(\bm{\mathcal{Y}}, \bm{\mathcal{X}}(\theta)), \label{eq:generative}
\end{align}
where $\bm{\mathcal{Y}}$ is a given signal, $\bm{\mathcal{X}}(\theta)$ is a parametric signal model with parameter $\theta$, and $D(\cdot,\cdot)$ is a distance measure between two inputs.
In the constrained approach, the signals/functions are directly constrained by the penalty function.
Typical examples are quadratic variation (QV) and total variation (TV) minimization \cite{rudin1992nonlinear}.
In the formula, it can be written as
\begin{align}
    \mathop{\text{minimize}}_{\bm{\mathcal{X}}} D(\bm{\mathcal{Y}}, \bm{\mathcal{X}}) + \lambda P(\bm{\mathcal{X}}), \label{eq:penalty}
\end{align}
where $P(\cdot)$ is a penalty function and $\lambda$ is a trade off parameter to balance $D$ and $P$.
From equations \eqref{eq:generative} and \eqref{eq:penalty}, both approaches do not compete.
In the formula, the hybrid approach is given as follows:
\begin{align}
    \mathop{\text{minimize}}_{\theta} D(\bm{\mathcal{Y}}, \bm{\mathcal{X}}(\theta)) + \lambda \tilde{P}(\theta), \label{eq:hybrid}
\end{align}
where we put $\tilde{P}(\theta) := P(\bm{\mathcal{X}}(\theta))$.
This includes regularized least square regression \cite{tibshirani1996regression}, sparse modeling \cite{elad2010sparse}, penalized matrix/tensor factorization models \cite{zdunek2007gibbs,zdunek2008blind,zdunek2012approximation,essid2013smooth,zdunek2014b,yokota2015smooth,yokota2016smooth,yokota2016tensor}.

In recent years, a method using delay embedding transformation has been attracted attention as a technique for reconstructing smooth signals \cite{brunton2017chaos}.
Nonetheless, it has provided the significant progress in combined with the recent technologies in computer science such as manifold learning \cite{erem2016extensions}, matrix/tensor factorization \cite{erem2016dynamic,yokota2018missing,yokota2018tensor}, time-series model \cite{shi2020block}, and neural networks \cite{yokota2019manifold}.
In \cite{yokota2018missing,yokota2018tensor}, a higher-order extension of delay embedding transform has been proposed, and combined with Tucker decomposition in higher-order delay-embedded space.
In \cite{sedighin2020matrix,sedighin2021image}, Sedighin \textit{et al.} proposed methods of applying tensor train and ring decompositions to a higher-order tensor with multistage delay embedding for signal restoration.
In \cite{shi2020block}, Shi \textit{et al.} proposed a method of applying auto-regressive integrated moving average (ARIMA) model in delay-embedded space for multiple short time-series forecasting.
In \cite{yokota2019manifold}, an image restoration technique by combining higher-order delay embedding transform and denoising auto-encoder has been proposed.  It also suggested that this has a close relationship with the image prior included in the convolutional neural network structure \cite{ulyanov2018deep}.
The property of the methods using delay-embedding is to capture the shift-invariant feature of signals.
It enables to perform smooth signal reconstruction by low-rank or low-dimensional approximation without using explicit smoothness constraint.

In this paper, we introduce a new aspect of signal restoration using delay embedding, which is different from the above studies.
We connect the convolutional model and the QV regularization through the delay-embedded space.
This paper contains ideas that may be useful in understanding and controlling the behavior of convolutional neural networks that have been actively studied in recent years.
The contributions of this study can be summarized as follows.
(1) We propose a new signal restoration model using smooth rank 1 matrix factorization and inverse delay embedding. The proposed model can be characterized as a generalization of QV regularization.
(2) It is shown theoretically and experimentally that the proposed model can realize soft smoothing compared to QV regularization.
This property plays important role in the problem of signal declipping.
(3) We show the properties of the proposed method in extensive experiments such as optimization behavior based on initial values, behavior change due to hyperparameters, basic behavior of signal reconstruction such as in signal declipping and noise removal, and comparison of signal recovery accuracy with real audio data.

\section{Delay-embedding}\label{sec:delay_embedding}

\subsection{Definition of delay-embedding}
Delay-embedding is defined as a linear operator inputting a vector $\bm y \in \bbR{N}$ and outputting a Hankel matrix $\bm Y \in \bbR{T \times \tau}$.
Each row of $\bm Y$ is identical to the local window of a vector $\bm y$.
For example, let us put $\bm y = [y_1, y_2, ..., y_5]^\top$ and $\tau=3$, its delay-embedded matrix can be given as
\begin{align}
    \bm Y := \mathcal{H}_\tau(\bm y) =
    \begin{pmatrix}
     c   & c   & y_1  \\
     c   & y_1 & y_2  \\
     y_1 & y_2 & y_3  \\
     y_2 & y_3 & y_4  \\ 
     y_3 & y_4 & y_5  \\ 
     y_4 & y_5 & c  \\ 
     y_5 & c   & c  
    \end{pmatrix},
\end{align}
where $c$ stands for the arbitrary number.
There are several options to decide the values of $c$.
Some padding operation (\eg zero-padding, reflection padding) is useful or to define $c$ as missing entries is one option.
In this study, we do not explicitly use this forward embedding then it is not necessary to define $c$ \footnote{Dare we say, $c$ can be seen as missing values in this study because these entries are not affected to objective function. }.
The column length of Hankel matrix $\bm Y$ is $T = N +\tau-1$ in this study.
In some study, there is a case to remove top and bottom two (\ie $\tau-1$) rows, and $T = N - \tau + 1$ in that case.

The pseudo inverse operation of delay-embedding can be defined as the average of anti-diagonal entries.
Let us put $Y(i,j)$ as the $(i,j)$-th entry of $\bm Y \in \bbR{T \times \tau}$, each entry of the inverse delay embedding can be given by
\begin{align}
    [\mathcal{H}_\tau^\dagger(\bm Y)](n) := \frac{1}{\tau} \sum_{t=1}^\tau Y( n + \tau - t , t). \label{eq:def_inv}
\end{align}
Note that the values of $c$ are not used in inverse operation.
In addition, the composition of forward and inverse embedding $\mathcal{H}_\tau^\dagger \circ \mathcal{H}_\tau$ is an identity mapping, however the opposite $\mathcal{H}_\tau \circ \mathcal{H}_\tau^\dagger$ is not when its input is not Hankel matrix.
By using a sparse matrix $\bm S_n \in \{0, \frac{1}{\tau} \}^{T \times \tau}$, inverse delay embedding can be rewritten by
\begin{align}
    [\mathcal{H}_\tau^\dagger(\bm Y)](n) = \langle \bm S_n, \bm Y \rangle.
\end{align}
Each entry of $\bm S_n$ is given by
\begin{align}
    S_n(i,j) = \left\{
    \begin{array}{l l}
        \frac{1}{\tau} & i = n + \tau - j \\
        0              & \text{otherwise}
    \end{array}\right. ,
\end{align}
Let us define a third order tensor $\mathcal{S}(n,i,j) := S_n(i,j)$, inverse delay embedding can be also rewritten by
\begin{align}
    \mathcal{H}_\tau^\dagger(\bm Y) = \bm S_{(1)} \text{vec}(\bm Y), \label{eq:inverse_delay_vec}
\end{align}
where a matrix $\bm S_{(1)} \in \{0, \frac{1}{\tau}\}^{N \times T\tau} $ is the mode-1 unfolding of a tensor $\ten{S} \in \{0, \frac{1}{\tau}\}^{N \times T \times \tau}$, and a vector $\text{vec}(\bm Y) \in \bbR{T\tau}$ is the unfolding of a matrix $\bm Y \in \bbR{T \times \tau}$.

\subsection{Low-rank matrix recovery in delay-embedded space}
In this section, we review the existing signal reconstruction method using low-rank model in embedded space \cite{yokota2018missing}.
This framework consists of the following three steps: (i) delay-embedding of a corrupted signal $\bm y$ to obtain $\bm Y$, (ii) obtaining $\bm X$ by using low-rank model, (iii) finally reconstructing the recovered signal $\bm x = \mathcal{H}_\tau^\dagger(\bm X) $.
The second step can be expressed as the following optimization problem
\begin{align}
   \mathop{\text{minimize}}_{\bm X} || P_\Omega (\bm Y - \bm X) ||_F^2,
                        \text{ s.t. } \bm X = \bm A \bm B^\top \label{eq:previous_model}
\end{align}
where $\bm A \in \bbR{T \times R}$ and $\bm B \in \bbR{\tau \times R}$.
Since $\bm Y$ is an incomplete matrix, Euclid distance between $\bm Y$ and $\bm X$ is calculated with support projection $P_\Omega$ which multiplies 1 to observed entries and 0 to unobserved entries.
Note that $\bm Y$ is a Hankel matrix partially and its low-rank approximation $\bm X$ is also automatically approaching to Hankel matrix partially.

\section{Discussions on a signal model with inverse delay-embedding}\label{sec:main_study}
The previous formulation Eq.\eqref{eq:previous_model} implicitly constrain the low-rank matrix $\bm X$ in delay-embedded space to be a Hankel matrix partially.
Here, we consider to remove this Hankel constraint from the model.
Differ from Eq.\eqref{eq:previous_model}, we do not explicitly use forward delay-embedding.
We directly define the reconstructed signal by $\bm x :=\mathcal{H}_\tau^\dagger(\bm X)$, and consider the model of a latent matrix $\bm X$.
The reduction of Hankel constraint provides the following properties to the model:
\begin{enumerate}
\item characterization as convolutional model (see Section~\ref{sec:convolution}),
\item sufficient representation ability even with a rank-1 matrix (see Section~\ref{sec:lower_rank}).
\end{enumerate}
Details of individual properties are explained later.

\subsection{Linear model with convolutional bases}\label{sec:convolution}
Any matrix $\bm X \in \bbR{T \times \tau} \ (\tau \leq T)$ has its singular value decomposition
\begin{align}
    \bm X = \bm U \bm \Sigma \bm V^\top = \sum_{r=1}^\tau \sigma_r \bm u_r \bm v_r^\top,
\end{align}
where $\bm U = [\bm u_1, ..., \bm u_\tau] \in \bbR{T \times \tau}$ and $\bm V = [\bm v_1, ..., \bm v_\tau] \in \bbR{\tau \times \tau}$ are orthonormal matrices which satisfy respectively $\bm U^\top \bm U = \bm I_\tau$ and $\bm V^\top\bm V = \bm I_\tau$, and $\bm \Sigma = \text{diag}(\sigma_1, ..., \sigma_\tau) \in \bbR{\tau \times \tau}$ is a diagonal matrix.

Next, we consider the inverse delay-embedding of $\bm X$.
Since the inverse delay-embedding is a linear operation, it can be separated into each rank-1 subspace,
\begin{align}
    \mathcal{H}_\tau^\dagger (\bm X) = \sum_{r=1}^\tau \sigma_r \mathcal{H}_\tau^\dagger(\bm u_r \bm v_r^\top).
\end{align}
Thus, it is a linear combination of inverse delay-embedding of rank-1 matrix.
From the definition of inverse delay-embedding \eqref{eq:def_inv}, we have
\begin{align}
    [\mathcal{H}_\tau^\dagger(\bm u_r \bm v_r^\top)](n) &= \frac{1}{\tau} \sum_{t=1}^\tau u_r(n+\tau-t) v_r(t) \notag \\
                                                     &= \frac{1}{\tau}[\bm v_r * \bm u_r](n).
\end{align}
In fact, inverse delay-embedding of rank-1 matrix is the convolution between two factor vectors.

Let us put a vector of singular values by $\bm \sigma = [\sigma_1, ..., \sigma_\tau]^\top$, vectorization of $\bm X$ and its inverse delay embedding can be rewritten by
\begin{align}
    \text{vec}(\bm X) &= (\bm V \odot \bm U) \bm \sigma, \label{eq:svd_vec} \\
    \tau\mathcal{H}_\tau^\dagger (\bm X) &= (\bm V \circledast \bm U)\bm \sigma,
\end{align}
where $\odot$ is Khatri-Rao product (\ie vector-wise Kronecker product), and $\circledast$ stands for the operation of vector-wise convolution.

From above derivations, we can understand that the inverse delay embedding of a latent matrix is a linear model with convolutional bases.  Its structure of vector-wise convolution and summation is actually equivalent to the convolutional layer in convolutional neural networks (CNN) \cite{fukushima1980neocognitron}.  For example, a color-image (or multi-channel image) represented by a matrix $\bm U \in \bbR{HW \times C_\text{in}}$, where $H, W, C_\text{in}$ are respectively height, width, and number of channels, can be assumed as an input of 2d-convolutional layer (2d-conv-layer) with convolutional kernel represented by matrices $\bm V_c \in \bbR{K^2 \times C_\text{in}}$ for all $c \in \{1, 2, ..., C_\text{out}\}$.  The CNN actually do the 2d-convolution along image domain (\ie $HW$ and $K^2$), and the summation along channel domain,
\begin{align}
   \text{2d-conv-layer}(\bm U, \{ \bm V_1, ..., \bm V_{C_\text{out}} \}) = [(\bm V_1 \circledast_\text{2d} \bm U)\bm 1, ..., (\bm V_{C_\text{out}} \circledast_\text{2d} \bm U)\bm 1] \in \bbR{HW \times C_\text{out}}
\end{align}
where $\circledast_\text{2d}$ is a operator of 2d-convolution, and $\bm 1 = [1, ..., 1]^\top$ is a $C_\text{in}$-dimensional vector of ones.

\begin{figure}
    \centering
    \includegraphics[width=0.69\textwidth]{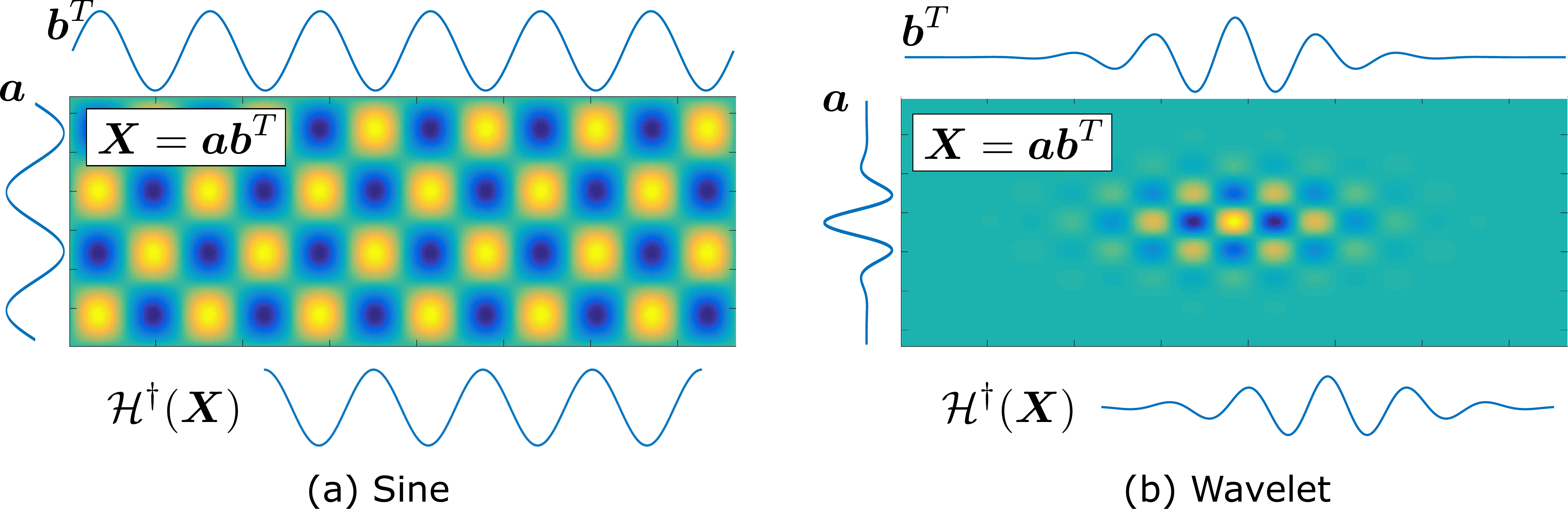}
    \caption{Rank-1 representation in latent space of sine and wavelet functions.}
    \label{fig:rank-1_sin_wavelet}
\end{figure}

\subsection{Sufficient representation ability even with a rank-1 matrix}\label{sec:lower_rank}
Here, we discuss the low-rank representation of $\bm X$.
From above derivation, rank of $\bm X$ is the number of convolutional bases.
Degree of freedom of each convolutional basis decides the representation ability of the model.

In this paper, we consider $\bm X$ as rank-1 matrix, and show it has sufficient representation ability as a generative model for signal reconstruction.
First, some natural signals like sine function can be generated by rank-1 matrix $\bm X = \bm a \bm b^\top$ with natural factor vectors $\bm a$ and $\bm b$.
We show two examples of sine and wavelet functions in Figure~\ref{fig:rank-1_sin_wavelet}.
Both signals are generated from rank-1 matrices.
In fact, convolution of sine functions is also sine function, and also convolution of two Gaussian is a Gaussian.

Actually, rank-1 matrix model of $\bm X = \bm a \bm b^\top \in \bbR{(N+\tau-1) \times \tau}$ can generate any $\bm x \in \bbR{N}$.
Let us put
\begin{align}
  \bm a = \begin{bmatrix} \bm 0_{\tau - 1} \\ \bm x \end{bmatrix} \in \bbR{N + \tau -1} \text{ and }
  \bm b = \begin{bmatrix} \tau \\ \bm 0_{\tau -1} \end{bmatrix} \in \bbR{\tau},
\end{align}
where $\bm 0_{\tau -1}$ is a $(\tau-1)$-dimensional vector of zeros, then we have
\begin{align}
  \mathcal{H}_\tau^\dagger(\bm a \bm b^\top) = \mathcal{H}_\tau^\dagger\left( \begin{bmatrix} \bm 0_{\tau-1} & \bm 0_{\tau-1,\tau-1} \\ \tau \bm x & \bm 0_{N,\tau-1} \end{bmatrix}  \right) = \bm x.
\end{align} 
This fact suggests us that the inverse-delay embedding of an unconstrained matrix even rank-1 is over-parameterized and it does not work as the model of smooth signals.
In this study, we consider to impose additional constraints to $\bm a$ and $\bm b$, and show some good properties of it for smooth signal reconstruction.

\section{Proposed signal model of soft smoothness}\label{sec:propose}
Here, we propose a new signal reconstruction model.
We assume the observed signal $\bm y \in \bbR{N}$ is incomplete that some entries have no values.
The diagonal projection matrix $\bm P_\Omega \in \{0, 1\}^{N \times N}$ passes observed entries and make missing entries to be zero.  The diagonal entries are given by
\begin{align}
    \bm P_\Omega(n,n) = \left\{ \begin{array}{ll} 1 & \text{if $y_n$ is observed ($n \in \Omega$)} \\
                                                    0 & \text{otherwise}
                                                    \end{array}\right. . \label{eq:projection}
\end{align}
The problem here is to obtain a complete signal $\bm x = \mathcal{H}_\tau^\dagger (\sigma\bm a \bm b^\top)$ from $\bm y$ by optimizing parameters $\bm a \in \bbR{T}$ and $\bm b \in \bbR{\tau}$.

In this paper, we impose smoothness constraint to $\bm a$ and $\bm b$.
Then, the optimization problem is given by
\begin{align}
    \mathop{\text{minimize}}_{\bm a, \bm b, \sigma} & \ \ || \bm P_\Omega (\bm y - \mathcal{H}_\tau^\dagger (\sigma \bm a \bm b^\top)) ||_2^2 
    + \lambda_a \sum_{i=1}^{T-1} (a_{i+1} - a_i)^2 + \lambda_b \sum_{j=1}^{\tau-1} (b_{j+1} - b_j)^2, \label{eq:optimization_problem}\\
    \text{s.t. } & \ \ || \bm a ||_2 = || \bm b ||_2 = 1, \notag
\end{align}
where $\sigma$ is a scalar variable, and $\lambda_a$, $\lambda_b$ are hyper-parameters.
The second and third terms stands for l2 penalty to provide smoothness, it is called as quadratic variation (QV) norm.
In addition, we impose unit vector constraint to $\bm a$ and $\bm b$ because QV penalties are affected by scales of $\bm a$ and $\bm b$.

\subsection{Properties}

\subsubsection{Characterization of generalized QV}
In this section, we show that QV regularization is a special case of the proposed method when $\tau=1$.

The QV regularization based signal reconstruction can be written as the following optimization problem:
\begin{align}
    \mathop{\text{minimize}}_{\bm x} || \bm P_\Omega (\bm y - \bm x) ||_2^2 + \lambda \sum_{i=1}^{N-1} (x_{i+1} - x_i)^2, \label{eq:QV_regularization}
\end{align}
where $\lambda$ stands for the level of QV regularization.
Small $\lambda$ performs weak smoothing, and large $\lambda$ performs strong smoothing.

On the other hand, when $\tau=1$ in the proposed model \eqref{eq:optimization_problem}, the inverse delay-embedding becomes an identity map and $\bm b$ is automatically determined to be 1.
Furthermore, when we put $\sigma \bm a = \bm x$ and $\lambda_a= \sigma^2 \lambda$, the proposed model \eqref{eq:optimization_problem} is reduced to \eqref{eq:QV_regularization}.

\subsubsection{Soft smoothing}
In this section, we analyse how does affect the smooth regularization of $\bm a$ and $\bm b$ to smoothness of the reconstructed signal $\bm x$.
To simplify the reconstruction model for the analysis, we consider to all $\bm x$, $\bm a$ and $\bm b$ are periodic functions and these sizes are the same.
Note that there would be a little gap between this analysis and real case because of the gap between a circular convolution and a linear (non-circular) convolution.

We assume $\bm x = \bm a * \bm b$ and its Fourier transform satisfies $\tilde{\bm x} = \tilde{\bm a} \circ \tilde{\bm b}$, where $\tilde{\cdot}$ stands for the Fourier transform and $\circ$ is entry-wise product.  Since differential operator is a convolution with a kernel $\bm l = [-1, 1, 0, ..., 0]^\top$, quadratic variation terms can be rewritten by
\begin{align}
    || \bm l * \bm a ||_2^2 + || \bm l * \bm b ||_2^2 &= || \tilde{\bm l} \circ \tilde{\bm a} ||_2^2 + || \tilde{\bm l} \circ \tilde{\bm b} ||_2^2 \notag \\
    &= \sum_{n=1}^N (\tilde{l}_n \tilde{a}_n)^2 + (\tilde{l}_n \tilde{b}_n)^2 \notag \\
    &= \sum_{n=1}^N \tilde{l}_n^2 (\tilde{a}_n^2 + \tilde{b}_n^2) \notag \\
    &\geq 2\sum_{n=1}^N \tilde{l}_n^2 | \tilde{a}_n \tilde{b}_n | \notag \\
    &= 2\sum_{n=1}^N \tilde{l}_n^2 | \tilde{x}_n |.
\end{align}
Here, we used inequality of arithmetic and geometric means.
Finally, we have
\begin{align}
    || \bm l * \bm a ||_2^2 + || \bm l * \bm b ||_2^2 \geq 2|| \tilde{\bm l} \circ \sqrt{| \tilde{\bm x} |} ||_2^2. \label{ineq:soft_smoothness}
\end{align}
The minimization of left-hand provides the minimization of right-hand, too.
Eq.\eqref{ineq:soft_smoothness} means that the smoothing both $\bm a$ and $\bm b$ make $\bm x$ to be smooth.  However, the levels of smoothing for individual frequencies are re-scaled as its squared root.
Since squared root scale down the higher values, the proposed signal reconstruction method may provide the softer smoothness than the simple quadratic variation regularization $ || \bm l * \bm x ||_2^2 $.
This property is also experimentally shown in Figures~\ref{fig:toy_exp_denoise} and \ref{fig:generalization_of_QV}.

\subsection{Optimization algorithm}
Here, we derive an optimization algorithm to solve \eqref{eq:optimization_problem}.
To improve the linear perspective of mathematical formulations, we put the objective function using matrices and a tensor as follow:
\begin{align}
     \mathcal{L}(\bm a, \bm b, \sigma) :=& || \bm P_\Omega (\bm y - \sigma \bm{\mathcal{S}} \times_2 \bm a^\top \times_3 \bm b^\top)) ||_2^2 + \lambda_a || \bm L_a \bm a ||_2^2 + \lambda_b || \bm L_b \bm b ||_2^2, \label{eq:objective_function}
\end{align}
where $\bm S \in \{0, \frac{1}{\tau} \}^{N \times T \times \tau}$ is the same tensor used in Eq.~\eqref{eq:inverse_delay_vec}, and $\bm L_a \in \{-1, 0, 1\}^{(T-1) \times T}$ and $\bm L_b \in \{-1, 0 ,1\}^{(\tau - 1) \times \tau}$ are differential operators of $\bm a$ and $\bm b$.
Both $\times_2$ and $\times_3$ stand for tensor-matrix product with mode-2 and mode-3, respectively.

An optimal condition of this problem can be given by
\begin{align}
    &\frac{\partial{\mathcal L}}{\partial\bm a} = 0, \frac{\partial{\mathcal L}}{\partial\bm b} = 0, \frac{\partial{\mathcal L}}{\partial\sigma} = 0, || \bm a ||_2 = || \bm b ||_2 = 1.
\end{align}
We solve the optimization problem by using alternating least squares (ALS) with projections shown in Algorithm~\ref{alg}.
Update rules for individual variables are follows.
For updating $\bm a$, we solve
\begin{align}
    [\sigma^2\bm S_{(2)}(\bm b \bm b^\top \otimes \bm I_{T})\bm S_{(2)}^\top + \lambda_a \bm L_a^\top \bm L_a] \bm a = \sigma \bm S_{(2)}(\bm b \otimes \bm P_{\Omega}\bm y), \label{eq:sub_a}
\end{align}
where $\bm I_{T}$ is an identity matrix with size of $(T,T)$, and $\otimes$ is Kronecker product.
After that we project $\bm a$ into unit ball.
For updating $\bm b$, we solve
\begin{align}
    [\sigma^2\bm S_{(3)}(\bm a \bm a^\top \otimes \bm I_{\tau})\bm S_{(3)}^\top + \lambda_b \bm L_b^\top \bm L_b] \bm b = \sigma \bm S_{(3)}(\bm a \otimes \bm P_{\Omega}\bm y), \label{eq:sub_b}
\end{align}
where $\bm I_{\tau}$ is an identity matrix with size of $(\tau,\tau)$.
After that we project $\bm b$ into unit ball.
For updating $\sigma$, we use the following least square solution:
\begin{align}
  \sigma \leftarrow \frac{\bm y^\top \bm P_\Omega \bm S_{(1)}(\bm b \otimes \bm a) }{|| \bm P_\Omega \bm S_{(1)}(\bm b \otimes \bm a) ||^2_2}.\label{eq:update_sigma}
\end{align}
Thus, we iterate to solve linear equations \eqref{eq:sub_a} with respect to $\bm a$, and \eqref{eq:sub_b} with respect to $\bm b$, and project $\bm a$ and $\bm b$ into unit ball, alternatively, until convergence.
Conjugate gradient method is recommended to solve each linear equation for efficient computation.

\begin{algorithm}[t]
\caption{ALS algorithm}\label{alg}
\begin{algorithmic}[1]
    \STATE {\bf Input}: $\bm y$, $\Omega$, $\tau$, $\lambda_a$, $\lambda_b$
    \STATE {\bf Initialize}: $\bm a$, $\bm b$, $\sigma$
    \REPEAT
        \STATE Update $\bm a$ by solving linear equation~\eqref{eq:sub_a};
        \STATE $\bm a \leftarrow \frac{\bm a}{||\bm a||_2}$;
        \STATE Update $\sigma$ by \eqref{eq:update_sigma};
        \STATE Update $\bm b$ by solving linear equation~\eqref{eq:sub_b};
        \STATE $\bm b \leftarrow \frac{\bm b}{||\bm b||_2}$;
        \STATE Update $\sigma$ by \eqref{eq:update_sigma};
    \UNTIL{convergence}
    \STATE {\bf Output}: $\bm a$, $\bm b$, $\sigma$
\end{algorithmic}
\end{algorithm}

\begin{algorithm}[t]
\caption{Outer algorithm}\label{alg:mc}
\begin{algorithmic}[1]
    \STATE {\bf Input}: $\bm y$, $\Omega$, $\tau$, $\lambda_a$, $\lambda_b$, $K$
    \FOR{$k = 1, 2, ..., K$}
        \STATE Initialize ($\bm a_k$, $\bm b_k$) randomly;
        \STATE $\bm a_k \leftarrow \frac{\bm a_k}{||\bm a_k||_2}$;
        \STATE $\bm b_k \leftarrow \frac{\bm b_k}{||\bm b_k||_2}$;
        \STATE Initialize $\sigma_k$ by \eqref{eq:update_sigma} using ($\bm a_k$, $\bm b_k$);
        \STATE Obtain $(\hat{\bm a}_k, \hat{\bm b}_k, \hat{\sigma}_k)$ by Algorithm~\ref{alg} with initialization $(\bm a_k, \bm b_k, \sigma_k)$;
    \ENDFOR
    \STATE $k^*  = \mathop{\text{argmin}}_{k} \mathcal{L}(\hat{\bm a}_k, \hat{\bm b}_k, \hat{\sigma}_k)$;
    \STATE {\bf Output}: $\bm a = \hat{\bm a}_{k^*}$, $\bm b = \hat{\bm b}_{k^*}$, $\sigma = \hat{\sigma}_{k^*}$
\end{algorithmic}
\end{algorithm}

\begin{figure}[t]
    \centering
    \includegraphics[width=0.99\textwidth]{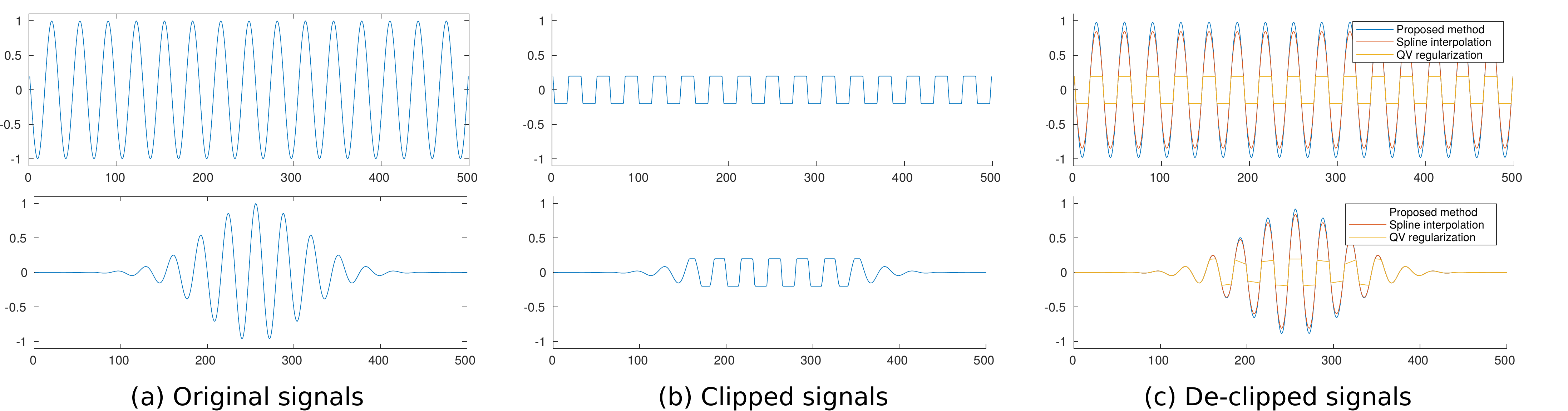}
    \vspace{-3mm}
    \caption{Examples of declipping experiments: (a) original signals of sine (top) and wavelet (bottom) functions, (b) clipped signals with clipping level = 0.2, and (c) these reconstructed signals by using QV regularization, cubic spline interpolation, and the proposed method.}\label{fig:toy_exp_plot}
    \vspace{3mm}
    \centering
    \includegraphics[width=0.65\textwidth]{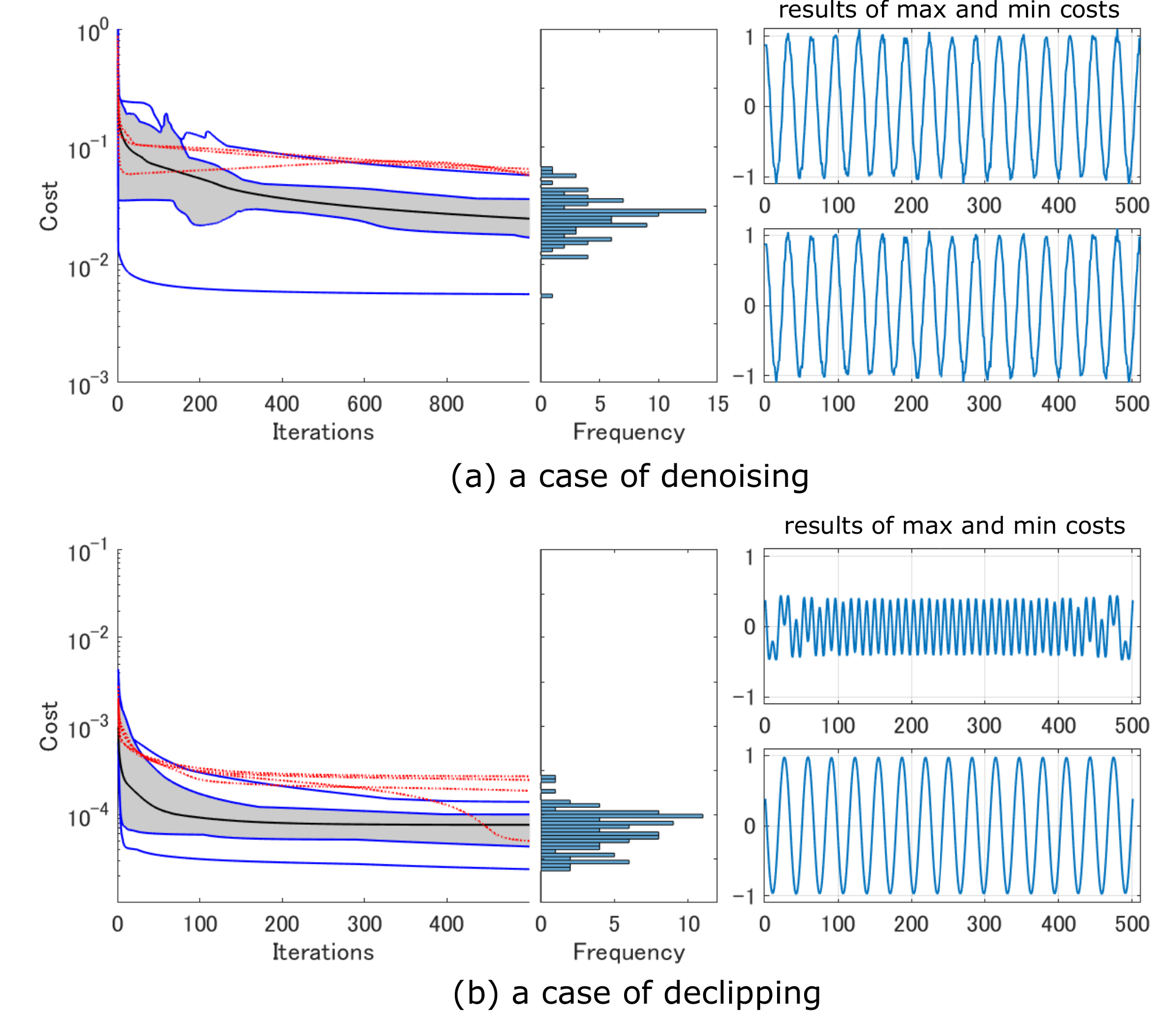}
    \vspace{-3mm}
    \caption{Optimization behavior: (a) results in denoising problem with additive Gaussian noise. (b) results in declipping problem with clipping level = 0.4. The left figure shows a functional boxplot of 100 curves for different 100 initial values.  The center figure shows a histogram of final 100 objective values.  The right figure shows two representative reconstructed signals with maximum (top) and minimum (bottom) objective values.}
    \label{fig:opt_behav}
\end{figure}

\subsubsection{Monte Carlo like outer algorithm to find better local optimum}
Since the non-convexity of the proposed optimization problem~\eqref{eq:optimization_problem}, the convergence point of $(\bm a, \bm b, \sigma)$ depends on the initialization.
To find better local optima, we propose to repeat $K$ times to run Algorithm~\ref{alg} with different random initialization of $\bm a$ and $\bm b$.
Then, we employ the minimum the best local optimum which performs the lowest value of objective function.
This outer Monte Carlo like method is summarized in Algorithm~\ref{alg:mc}

\section{Experiments}\label{sec:exp}

\subsection{Reconstruction of typical signals}
In this section, sine and wavelet functions are used to show the optimization behaviors of the proposed method, the hyperparameter sensitivities, and the qualitative differences from other methods.

There are several types of signal corruption.
Here, we consider additive noise and clipping.
Additive noise model is given by
\begin{align}
    \bm y = \bm y_0 + \bm e,
\end{align}
where $\bm y_0 \in \bbR{N}$ is an original signal and $\bm e \in \bbR{N}$ is noise.
We assume each entry $e(n)$ is independently sampled from an identical Gaussian distribution.

Clipping is an operation that uses a certain clipping level $c > 0$ to replace entries above $c$ and entries below $-c$ with $c$ or $-c$.
The clipping operation can be given by the following equation: 
\begin{align}
  \bm y = \min(c, \max(-c,\bm y_0)). \label{eq:clipping}
\end{align}
The value range of the clipped signal is $[-c,c]$.
Examples of declipping are shown in Figure~\ref{fig:toy_exp_plot}.
The original signal is shown in (a) and the signal after clipping is shown in (b).
The indices of the clipped entries are recorded and treated as missing values to be restored in this study.
Thus, the set of remained entries can be given as $\Omega := \{ n |  -c \leq y_0(n) \leq c \}$ and $\bm P_\Omega$ can be defined by \eqref{eq:projection} to apply the proposed method.

\begin{figure}[t]
    \centering
    \includegraphics[width=0.99\textwidth]{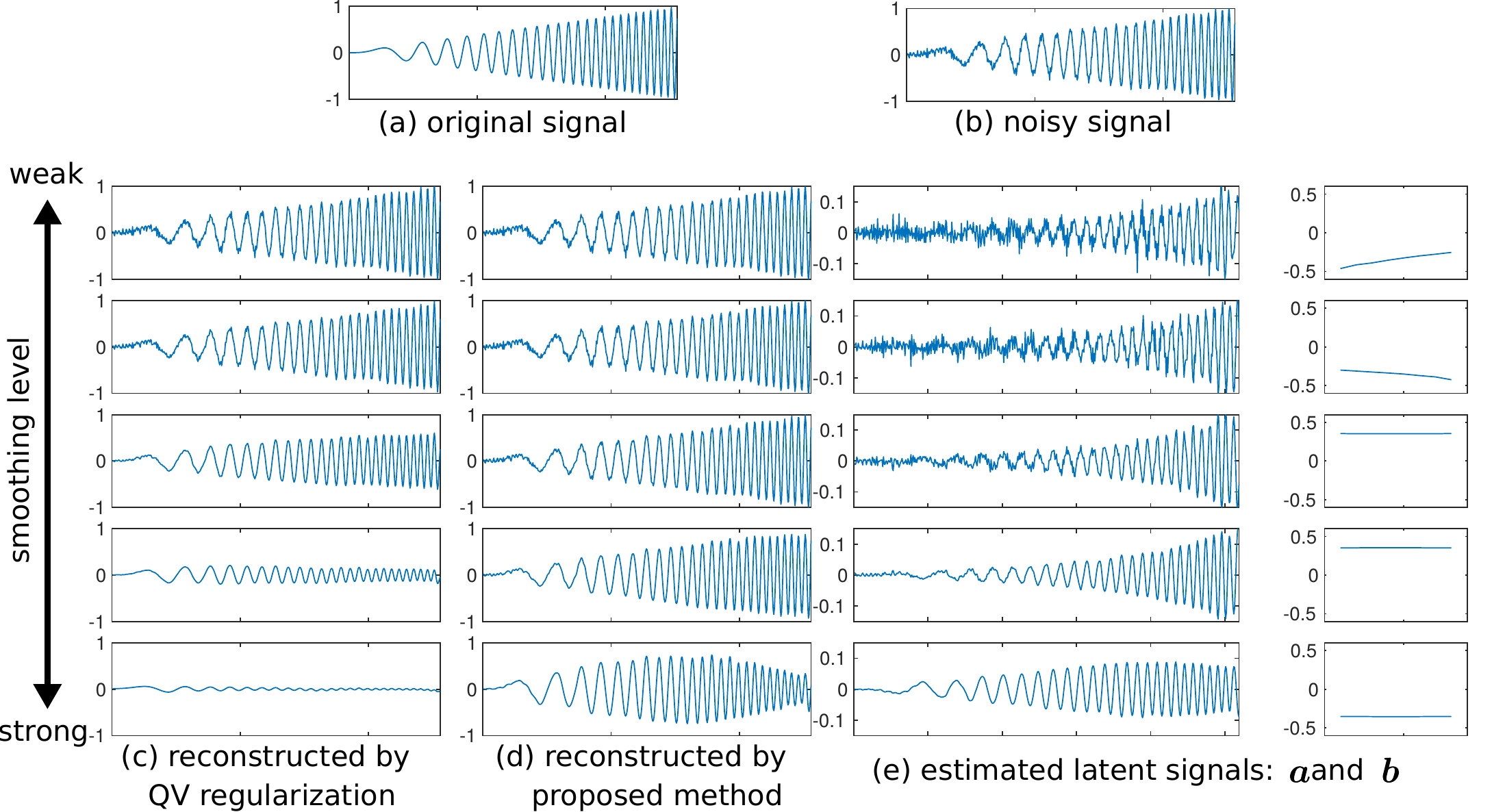}
    \caption{Reconstruction of noisy chirp signal by QV regularization and the proposed method: (a) original chirp signal, (b) noisy chirp signal, (c) reconstruction results by QV regularization, (d) reconstruction results by the proposed method, and (e) estimated latent signals $\bm a$ and $\bm b$.}\label{fig:toy_exp_denoise}
\end{figure}

\begin{figure}[t]
\centering
\includegraphics[width=0.69\textwidth]{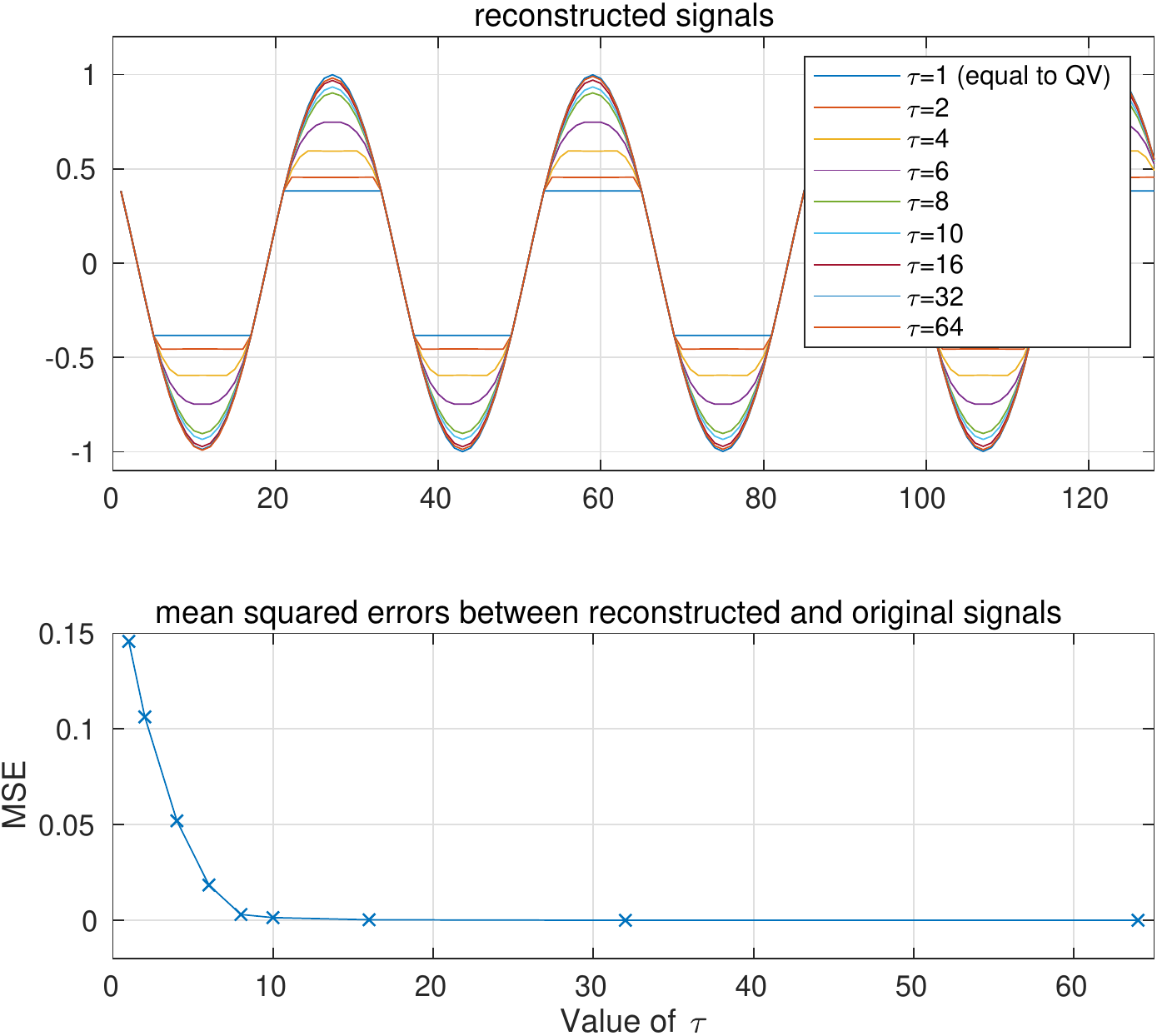}
\caption{Reconstructed signals of clipped sine curve ($c=0.4$) with various values of $\tau$.}\label{fig:generalization_of_QV}
\end{figure}

\subsubsection{Optimization behavior}
The optimization problem \eqref{eq:optimization_problem} is non-convex.
Therefore, the obtained results vary depending on the initial value.
In this section, we show the optimization behavior of proposed algorithm by randomly generating the initial values of $\bm a$ and $\bm b$.
Note that the initial value of $\sigma$ is determined by \eqref{eq:update_sigma} using initialization of ($\bm a$, $\bm b$).

Figure~\ref{fig:opt_behav} shows the changes in the objective function with randomly generated initial values, the histogram of the final values of objective function, and the representative reconstruction results.
One hundred of initial values of $(\bm a, \bm b)$ are randomly generated from the Gaussian distribution.
In each case, the parameters were updated 1000 or 500 times, which yielded one hundred curves showing the change in the objective function.
These curves were visualized using representative curves, the center line, the 25th percentile line, the 75th percentile line, the upper bound, the lower bound, and the outlier lines by functional boxplot \cite{sun2011functional}.
The histograms of a hundred objective function values after 1000 or 500 updates were visualized.
Finally, the reconstructed signals were drawn for each case where the objective function value was the maximum and minimum.

Figure~\ref{fig:opt_behav} (a) shows the optimization behavior of the proposed method in denoising problem.
A noisy sine function was restored by the proposed method.
Convergent points were varied with different initialization.
However, reconstructed signals are quite similar between two cases of maximum and minimum objective values.
Figure~\ref{fig:opt_behav} (b) shows the optimization behavior of the proposed method in declipping problem.
Components of sine function more than 0.4 and less than -0.4 were clipped, and its incomplete signal was recovered by the proposed method.
In this case too, the solutions varied depending on the initial values.
However unlike the denoising problem, reconstructed signals are quite different between two cases of maximum and minimum objective values.

\subsubsection{Smoothing behavior in compared with quadratic variation}
In this section, we demonstrate how the proposed method smooths the signal in comparison with quadratic variation (QV) regularization.
Figure~\ref{fig:toy_exp_denoise} (c) shows the result that a noisy chirp signal is reconstructed by QV regularization with five levels of lambda values.
If the regularization is too weak, the noise will not be removed sufficiently.
Conversely, if the regularization is too strong, not only the noise but also the original signal components will be removed.
The difficulty of this problem is that the high-frequency component at the second half of chirp signal is originally large, and this is also suppressed by smoothing.

On the other hand, Figure~\ref{fig:toy_exp_denoise} (d) shows the results of the signal reconstruction using the proposed method.
This shows that the smoothing properties are different from those of QV.
The proposed method prevents over-smoothing unlike QV regularization.

\subsubsection{Softer smoothing than QV regularization}
Figure~\ref{fig:generalization_of_QV} shows the results of declipping problem of sine function for various values of $\tau$.
Sine function is clipped with $c=0.4$, and it is declipped by Algorithm~\ref{alg:mc} with $K=10$.
Here, $\lambda_a$ and $\lambda_b$ are controlled with $\lambda_1$ and $\lambda_2$ by
\begin{align}
    \lambda_a &= \lambda_1\frac{|\Omega|}{TN}, \\
    \lambda_b &= \lambda_2\frac{|\Omega|}{\tau N},
\end{align}
where $|\Omega|$ stands for the number of observed entries.
We varied $\tau \in \{1, 2, 4, 6, 8, 10, 16, 32, 64\}$, and smoothness parameter is fixed as $\lambda_1 = \lambda_2 = 0.1$.
Note that the proposed method with $\tau=1$ is equivalent to QV regularization.
The QV regularization does not restore the component that jumps up and down of sine function.
When $\tau$ is increased, we can see the amplitude of clipped part is recovered and mean squared errors (MSE) decreases.

\subsubsection{Hyper-parameter sensitivities}
The proposed method has three hyper-parameters: the embedded dimension $\tau$ and the smoothing level $\lambda_1$ and $\lambda_2$.
In this section, we show the experimental results with declipping in various values of hyper-parameters: $\tau$, $\lambda_1$ and $\lambda_2$.

We recovered clipped sine and wavelet functions with the clip level of 0.4.
Three hyper-parameters were varied in the range $\tau \in \{8, 16, 32, 64, 128, 256\}$ and $\lambda_1, \lambda_2 \in \{0.001, 0.01, 0.1, 1, 10\}$.
Declipping by Algorithm~\ref{alg:mc} with $K=10$ was performed for all combinations (\ie 150 ways) to calculate the recovery accuracy.

\begin{figure}[t]
    \centering
    \includegraphics[width=0.65\textwidth]{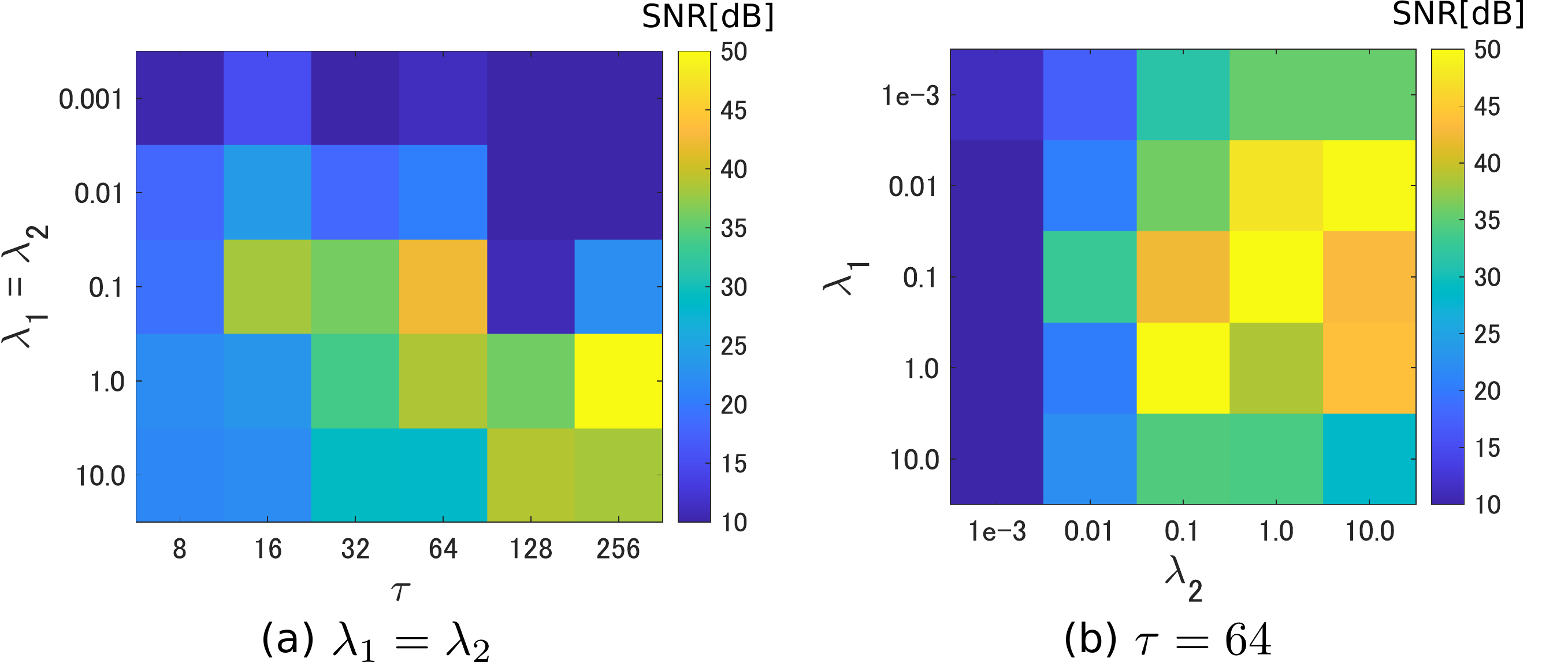}\vspace{-3mm}
    \caption{Effects of hyper-parameters in declipping sine function (clipping level = 0.4).  Reconstruction quality was evaluated by SNR for various settings of hyper-parameters $\lambda_1$, $\lambda_2$, and $\tau$.}\label{fig:var_para_sine}
\vspace{5mm}
    \centering
    \includegraphics[width=0.65\textwidth]{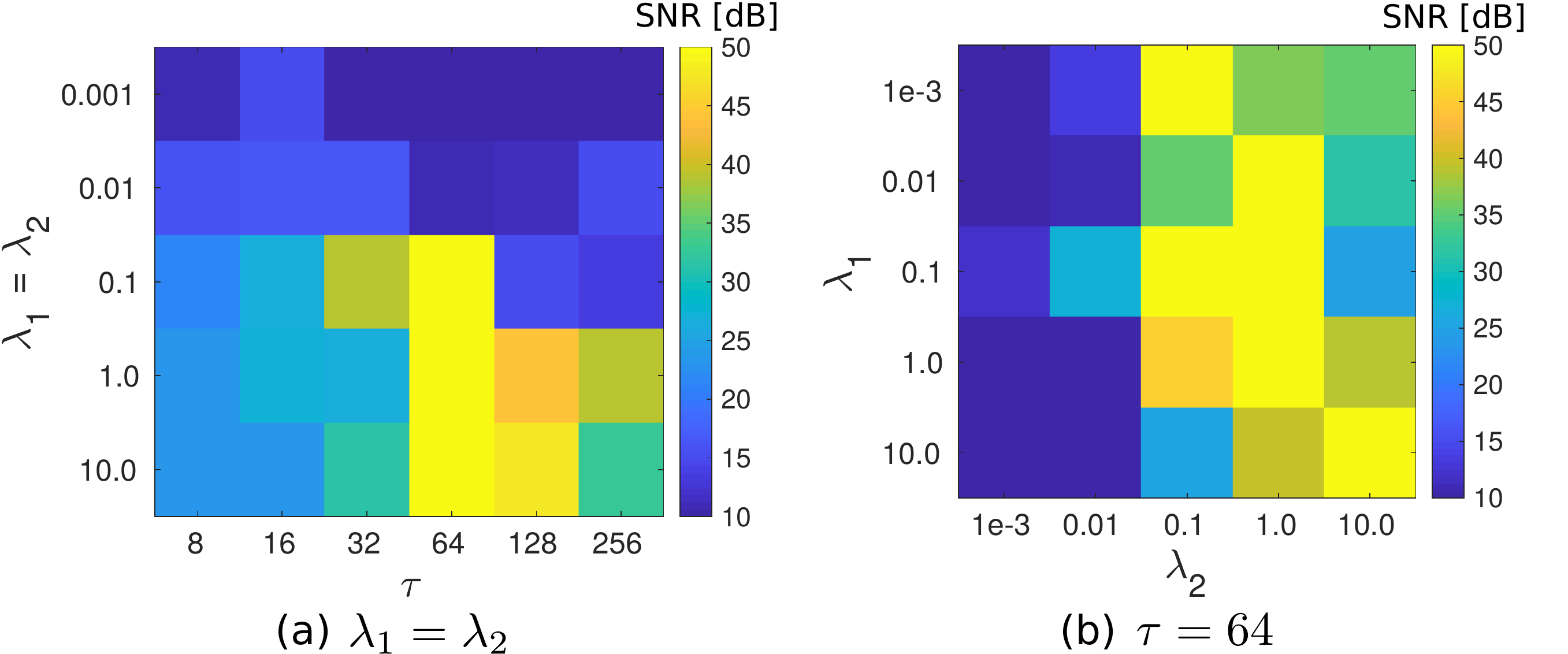}\vspace{-3mm}
    \caption{Effects of hyper-parameters in declipping wavelet function (clipping level = 0.4).  Reconstruction quality was evaluated by SNR for various settings of hyper-parameters $\lambda_1$, $\lambda_2$, and $\tau$.}\label{fig:var_para_wave}
    \vspace{5mm}
    \centering
    \includegraphics[width=0.79\textwidth]{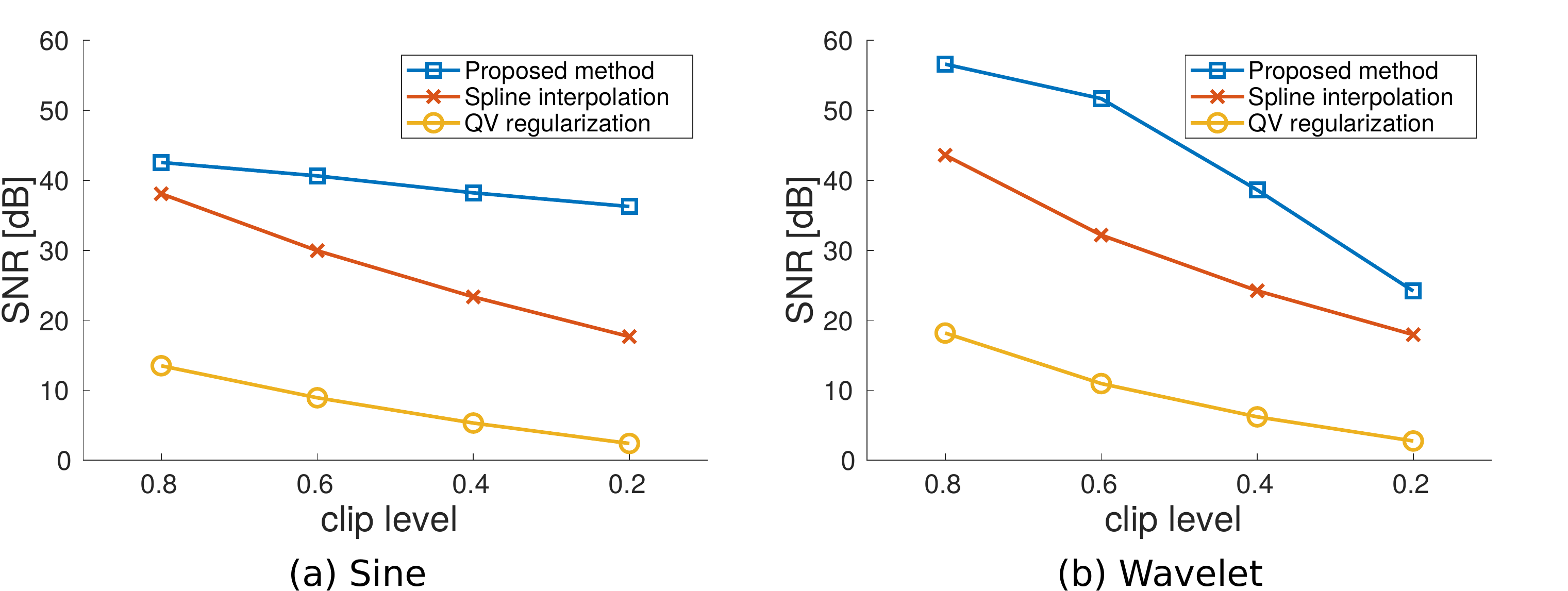}
    \vspace{-3mm}
    \caption{Values of SNR in declipping experiments with various clipping levels.}\label{fig:toy_exp_snr}
\end{figure}

The results are shown in Figures~\ref{fig:var_para_sine} and \ref{fig:var_para_wave}.
The change under the condition of $\lambda_1=\lambda_2$ is shown in (a).
It can be seen that the accuracy is better when $\tau$, $\lambda_1$ and $\lambda_2$ are relatively large.
Larger $(\lambda_1, \lambda_2)$ is preferred for larger $\tau$.
Next, we show in (b) the change under the condition of $\tau=64$, where we find that the accuracy is good for a wide range of $\lambda_1$ if $\lambda_2$ is large.
Almost same results are obtained for both sine and wavelet functions.

\begin{figure}[t]
    \centering
    \includegraphics[width=0.89\textwidth]{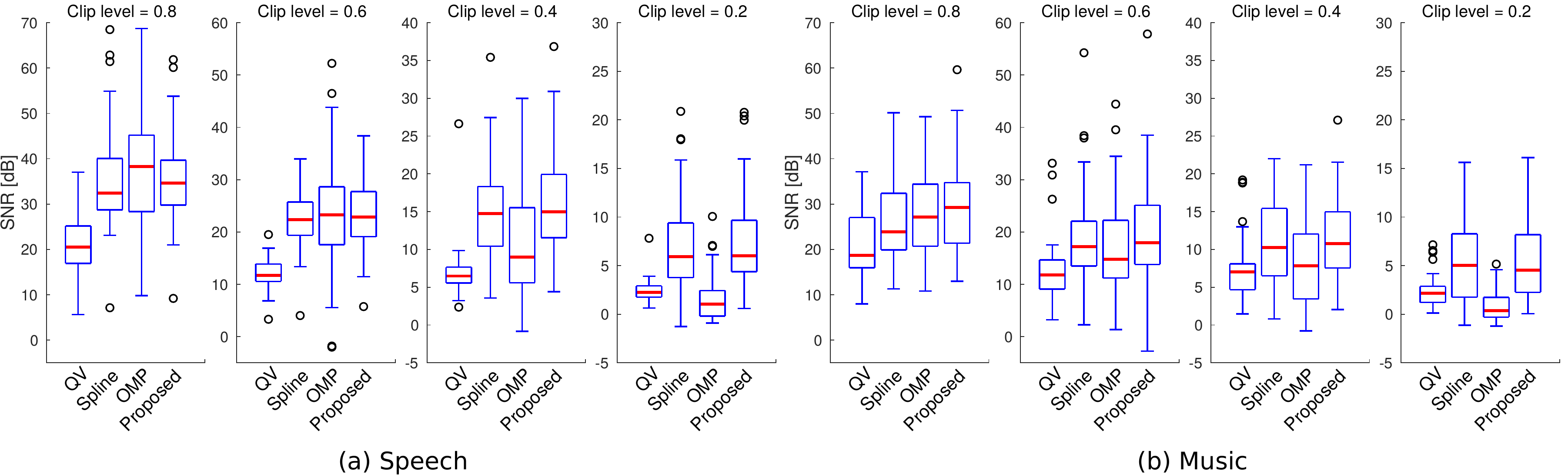}
    \caption{Box-plot of SNR values in declipping experiments with clip levels 0.8, 0.6, 0.4, and 0.2.  Segments of speech and music audio signals were recovered by quadratic variation regularization (QV), cubic spline interpolation (Spline), orthogonal matching pursuit (OMP), and the proposed method.}\label{fig:audio_declip}
\end{figure}

\begin{table}[t]
    \caption{Average and standard deviation of SNR in declipping}
    \label{tab:audio_depclip}
    \centering
    \begin{tabular}{c|c c c c}
    & QV & Spline & OMP & Proposed \\ \hline
Speech(0.8) & 21.4 $\pm$ 6.0 & 35.2 $\pm$ 9.8 & {\bf 37.7 $\pm$ 12.4} & 35.5 $\pm$ 8.7 \\ 
Speech(0.6) & 12.0 $\pm$ 2.6 & 22.2 $\pm$ 4.9 & {\bf 23.2 $\pm$ 10.0} & 23.1 $\pm$ 5.8 \\ 
Speech(0.4) & 6.7 $\pm$ 2.7 & 14.7 $\pm$ 5.4 & 10.4 $\pm$ 7.0 & {\bf 15.5 $\pm$ 5.9} \\ 
Speech(0.2) & 2.3 $\pm$ 1.0 & 6.8 $\pm$ 4.3 & 1.4 $\pm$ 2.1 & {\bf 7.3 $\pm$ 4.6} \\ \hline
Music(0.8)& 20.8 $\pm$ 7.0 & 26.6 $\pm$ 9.6 & 27.9 $\pm$ 8.9 & {\bf 29.6 $\pm$ 10.1} \\ 
Music(0.6)& 12.6 $\pm$ 5.2 & 18.2 $\pm$ 8.5 & 16.8 $\pm$ 8.2 & {\bf 19.3 $\pm$ 8.8} \\ 
Music(0.4)& 7.1 $\pm$ 3.6 & 11.0 $\pm$ 5.6 & 8.0 $\pm$ 5.4 & {\bf 11.6 $\pm$ 5.3} \\ 
Music(0.2)& 2.4 $\pm$ 1.5 & 4.4 $\pm$ 6.6 & 0.8 $\pm$ 1.4 & {\bf 5.4 $\pm$ 3.8} 
    \end{tabular}
\end{table}

\subsubsection{Declipping behavior in comparison with quadratic variation and spline interpolation}
In this section, we apply the proposed method to the declipping problem and compare it with QV regularization and spline interpolation.
In the proposed method, we set $\tau=64$ and $\lambda_1 = \lambda_2 = 1.0$.

Figure~\ref{fig:toy_exp_plot} (b) shows the clipped sine and wavelet signals with clipping level $c=0.2$.
These reconstructed signals by QV regularization, spline interpolation, and the proposed method are shown in Figure~\ref{fig:toy_exp_plot} (c).
The QV regularization shows that the missing entries are linearly interpolated and the components that jumps up and down are not restored at all.
In the spline interpolation, a third-order polynomial function is obtained so that the differential value at the boundary between the signal and the missing part is kept, and the smooth interpolation is performed.
In this example of the sine and wavelet functions, a signal is restored that seems to have a slightly smaller actual amplitude.
On the other hand, the proposed method recovered both sine and wavelet functions with high accuracy.

Figure~\ref{fig:toy_exp_snr} shows signal-to-noise ratio (SNR) values of this declipping experiments with various clipping levels.
The proposed method achieved significantly higher values of SNR than QV regularization and spline interpolation.

\subsection{Applications to audio inpainting}
In this section, we compare the proposed method with existing methods for audio inpainting \cite{adler2011audio,adler2011constrained}.

\subsubsection{Problem and data}
Audio inpainting is a method to estimate missing entries of a single audio signal.
In this study, we addressed two types of missing: clipping and random missing.
Four levels of clipping $c \in \{0.8, 0.6, 0.4, 0.2\}$ and four rates of missing $\{10\%, 30\%, 50\%, 70\%\}$ are tested.
Noise was not assumed.

We used 10 speech signals and 10 music signals which are packaged in audio inpainting toolbox\footnote{http://small.inria.fr/keyresults/audio-inpainting/}.
Sampling frequency of all signals is 16kHz.
Various segments of size 128 (approximately 0.01 seconds) with sufficient amplitude were extracted from these signals and used in the experiment.
Number of segments were respectively 78 and 59 from speech and music signals for clipping.
Number of segments were respectively 100 and 100 from speech and music signals for random missing.

\subsubsection{Methods in comparison}
In this experiments, we selected QV regularization, cubic-splines, and orthogonal matching pursuit (OMP) \cite{adler2011audio,adler2011constrained} as the methods in comparison.
These are representative signal reconstruction or interpolation methods.
QV regularization is defined in Eq.~\eqref{eq:QV_regularization}.
Cubic-splines perform smooth interpolation using segments of cubic curves that have the same gradient at the connection points.
OMP \cite{adler2011audio,adler2011constrained} approximately solves the following optimization problem: 
\begin{align}
    \mathop{\text{argmin}}_{\bm w} \ \  || \bm w ||_0, \text{s.t. } || \bm P_\Omega(\bm y - \bm D \bm w) ||_2^2 \leq \epsilon
\end{align}
where $|| \cdot ||_0$ is a $l_0$-norm which counts number of non-zero entries, $\bm D \in \bbR{N \times \rho N}$ is a redundant dictionary, $\bm w \in \bbR{\rho N}$ is a coefficient parameter, $\rho > 1$ is a level of redundancy, and $\epsilon$ is a small scalar.
We used Gabor dictionary with $\rho=2$, and set $\epsilon=0.001$.
Finally, we set $\tau=128$, $\lambda_1 = \lambda_2 = 1.0$, and $K=20$ in the proposed method.

\begin{figure}[t]
    \centering
    \includegraphics[width=0.99\textwidth]{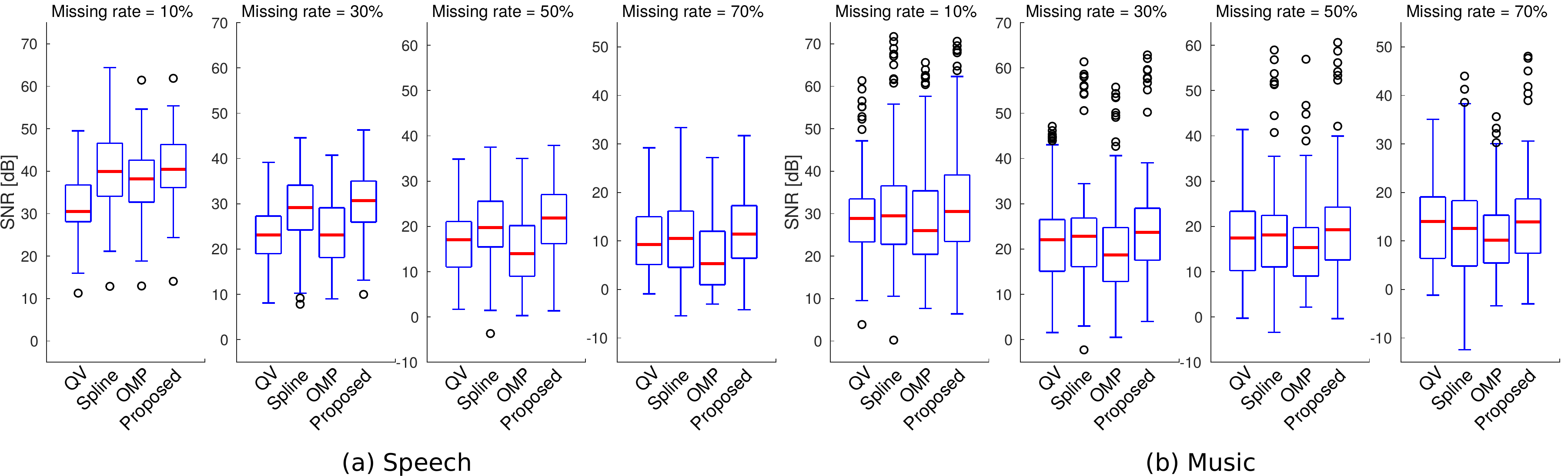}
    \caption{Box-plot of SNR values in completion experiments with random missing rates 10\%, 30\%, 50\%, and 70\%. Segments of speech and music audio signals were recovered by quadratic variation regularization (QV), cubic spline interpolation (Spline), orthogonal matching pursuit (OMP), and the proposed method.}\label{fig:audio_completion}
\end{figure}

\begin{table}[t]
    \caption{Average and standard deviation of SNR in completion}
    \label{tab:audio_completion}
    \centering
    \begin{tabular}{c|c c c c}
    & QV & Spline & OMP & Proposed \\ \hline
Speech(10\%)& 32.4 $\pm$ 7.3 & 39.8 $\pm$ 8.5 & 37.7 $\pm$ 8.0 & {\bf 40.7 $\pm$ 8.0} \\ 
Speech(30\%)& 23.1 $\pm$ 7.0 & 28.8 $\pm$ 7.5 & 23.5 $\pm$ 8.0 & {\bf 30.1 $\pm$ 7.2} \\ 
Speech(50\%)& 16.6 $\pm$ 6.8 & 20.0 $\pm$ 7.7 & 14.9 $\pm$ 7.7 & {\bf 21.3 $\pm$ 7.6} \\ 
Speech(70\%)& 10.4 $\pm$ 6.7 & 11.0 $\pm$ 8.4 & 7.0 $\pm$ 7.0 & {\bf 12.3 $\pm$ 8.1} \\ \hline
Music(10\%)& 29.3 $\pm$ 11.3 & 30.9 $\pm$ 14.4 & 29.7 $\pm$ 13.4 & {\bf 32.6 $\pm$ 13.9} \\ 
Music(30\%)& 22.1 $\pm$ 10.2 & 23.2 $\pm$ 13.4 & 20.9 $\pm$ 12.0 & {\bf 25.2 $\pm$ 13.4} \\ 
Music(50\%)& 17.8 $\pm$ 9.2 & 18.5 $\pm$ 12.5 & 15.9 $\pm$ 10.0 & {\bf 20.1 $\pm$ 12.2} \\ 
Music(70\%)& 13.4 $\pm$ 8.4 & 12.6 $\pm$ 10.7 & 10.9 $\pm$ 8.0 & {\bf 14.1 $\pm$ 10.7}
    \end{tabular}
\end{table}

\subsubsection{Results}
Figure~\ref{fig:audio_declip} and Table~\ref{tab:audio_depclip} show reconstruction accuracy (SNR) of audio declipping experiments.
The restoration accuracy decreased as the clip level decreased for both speech and music signals.
QV regularization was not accurate at all clip levels.
At the speech declipping with $c=0.8, 0.6$, OMP outperforms other methods, however the variance of SNR were slightly large.
Furthermore, the accuracy decrease of OMP with $c=0.4, 0.2$ was remarkable. 
On the other hand, spline and the proposed method were able to suppress the decrease in accuracy when the clip level was small. 
Especially for music signals, the average of SNR values of the proposed method was the highest at all clip levels.

Figure~\ref{fig:audio_completion} and Table~\ref{tab:audio_completion} show reconstruction accuracy (SNR) of audio completion experiments.
In contrast to declipping, QV regularization was competitive with other methods.
For instance, QV regularization outperforms OMP at the highly missing rates such as 50\% and 70\%.
Spline and the proposed methods were competitive with each other, but the proposed method was slightly better in terms of the average SNR at all missing rates.

\section{Conclusions}\label{sec:con}
In this paper, we proposed a new smooth signal model using constrained rank-1 matrix factorization with inverse delay-embedding.  It is characterized as a generalization of QV regularization, and performs softer smoothing than QV regularization at larger $\tau$.  We shown that this property helps us to recover the clipped signals and have a benefit in application of audio inpainting.
Future works may include the extension to image/tensor recovery and improvements of optimization algorithm.

\bibliographystyle{plain}

\end{document}